\documentclass[
amsmath, aps, prl 
]{revtex4}
\usepackage{bm}
\usepackage{graphicx}
\usepackage{amssymb}

\def \d{\partial}

\def \bv{{\bf v}}

\def \br{{\bf r}}

\def \bB{{\bf B}}
\def \bp{{\bf p}}

\def \br{{\bf r}}

\def \bphi{\boldsymbol{\phi}}

\begin{document}

\title{No feedback is possible in small-scale turbulent magnetic field}
\author{K.P. Zybin $^{1,2}$, A.S. Il'yn $^{1,2}$, A.V. Kopyev$^1$, and  V.A. Sirota $^{1}$\footnote{Electronic addresses: zybin@lpi.ru, asil72@mail.ru, kopyev@lpi.ru, sirota@lpi.ru}}

\affiliation 
{$^1$ P.N.Lebedev Physical Institute of RAS, 119991, Leninskij pr.53,
Moscow, Russia  \\
$^2$ National Research University Higher School of Economics, 101000,
Myasnitskaya 20, Moscow, Russia}

\begin{abstract}
Evolution of  stochastically  homogeneous magnetic field advected
by incompressible turbulent flow  with large magnetic Prandtl numbers
 is considered at the scales less than Kolmogorov viscous scale. It is shown that, despite unlimited growth of the magnetic field, its feedback on the fluid's dynamics remains negligibly small.
\end{abstract}

\maketitle

Generation of magnetic field in many physical and astrophysical systems, e.g., in planets, stars and galaxies, still remains unrevealed. Turbulent dynamo is the most natural and generally considered mechanism to increase the initial seed magnetic field in conductive media \cite{Moffat, KraichnanNagarajan, FGV}.
 The idea of the dynamo  is that the random transport stretches passively entrapped  magnetic lines, thus increasing fluctuations of magnetic field.
%
 In the case of large magnetic Prandtl numbers (e.g., in the interstellar medium),
dynamo is very efficient at the smallest scales
because of the exponential recession of trajectories of liquid particles. 
Magnetic field generation in this range has been a subject of many works with different approaches \cite{Kraichnan,Kulsrud92, Kazantsev,  zeld, Chertkov, Scripta18}
If the scales under consideration are smaller than Kolmogorov viscous scale $r_\eta$, one can use the approximation of linear velocity field, which corresponds to essential viscosity; this approximation is called Batchelor regime \cite{Batchelor}.
The case was considered in \cite{zeld, Chertkov, Scripta18}.
The classical problem statement of exponential growth of statistically homogeneous magnetic field fluctuations in this regime was considered in details [8]. It is general point that the exponential growth lasts until the feedback of magnetic field on velocity dynamics becomes important. In this article we show that in such statement the system never reaches the nonlinear saturation as it was previously believed.
%


Actually, what stops the unlimited exponential increase? The initial magnetic field fluctuations are supposed to be small.
This allows to consider the velocity dynamics independently of the magnetic field evolution, and makes the equation for magnetic flux density linear.
 The equation remains linear until the feedback of magnetic field on velocity dynamics becomes important. So it has been generally believed that this  feedback-caused nonlinearity   provides the restriction.

In the paper we study the details of the feedback. We consider the magnetohydrodynamic equations and analyze the Lorentz force that affects the velocity field. It can be divided into two terms: one of these terms has a gradient form and results in renormalization of  pressure, so it does not affect the velocity dynamics.
For the other term we show that, although it contains the second order function of the magnetic induction and despite of the exponential growth of magnetic field, this term decreases.
The reason is that stochastic magnetic field is not structureless: the regions of high induction have the form of tubes and sheets, and gradient of magnetic field is 
always almost exactly orthogonal to  the magnetic line.
So, its product with magnetic induction remains small and the feedback on Batchelor turbulence remains negligible, even though the induction itself increases.

Thus, we show that the feedback would not stop the dynamo mechanism and has no relation to the eventual value of magnetic field fluctuations in the viscous range of turbulence.
This raises the question of alternative restrictions. They may come
 from finite size 
of the region occupied by initial magnetic field \cite{zeld,epl18}, or,
in the homogenous limit,  from Kolmogorov scale: 
as lengthening magnetic lines reach the size 
$r_\eta$, their exponential stretching ceases. The characteristic time of exponential growth is then proportional to the logarithm of viscous-to correlation ($l$) scale ratio.

This degeneration of feedback might be a hint for even more complicated problems: something similar might take place in turbulence, giving possibility to find a key to the nonlinearity of the hydrodynamic equations.

\vspace{0.4cm}

The MHD equations describing joint evolution of velocity and magnetic fields in incompressible medium can be written as:
\begin{equation} \label{1-NS}
\frac{\partial \bv}{\partial t} + ({\bf v}  \nabla) \bv +\nabla p = \nu \Delta {\bf v} +\bphi + {\mathbf f}
\end{equation}
\begin{equation} \label{1}
\frac{\partial \bB}{\partial t} + ({\bf v} \nabla) \bB - (\bB  \nabla) {\bf v} = \varkappa \Delta \bB
\end{equation}
Here $\nu$ is viscosity of the fluid, $\varkappa$ is its magnetic diffusivity, $p$ is pressure; the $\bphi$ is a large-scale ($L$) pumping force 
to make the flow stationary.
Hereafter we assume that the magnetic diffusive scale $r_d$, the scale of initial magnetic field fluctuations $l$, and the viscous scale $r_{\eta}$ are related by
$$
r_d \le l \ll r_{\eta}  \ll L
$$
The Lorentz force $\bf f$ represents the feedback of magnetic field on the velocity dynamics,
\begin{equation} 
{\mathbf f} = - \tfrac 1{4\pi} \bB \times \left[ \nabla \times \bB \right]
\end{equation}
Both $\bB$ and $\bv$ are stochastic.

First, we note that $\mathbf f$ can be written as a sum of two terms:
\begin{equation} \label{force}  
{\mathbf f} =  \tfrac 1{4 \pi} (\bB \nabla) \bB - \tfrac 1{8\pi} \nabla (B^2)
\end{equation}
The second term is only a renormalization of pressure, it can be eliminated from Eq.(\ref{1-NS}) by changing the variable,
$$
p'=p + \tfrac 1{8\pi} B^2
$$
The only equation that governs pressure dynamics is the divergence of (\ref{1-NS}), i.e.,
\begin{equation} \label{for-pressure}
(\d_i v_j)(\d_j v_i) + \Delta p' = \tfrac{1}{4\pi}(\d_i B_j)(\d_j B_i)
\end{equation}
So, without any loss of information we can exclude the gradient term from (\ref{force}).

Second, if magnetic field is small enough, $\mathbf f$ is negligible (as well as the right-hand side of (\ref{for-pressure})),
and Eq.(\ref{1-NS}) is independent of $\bf B$. In the passive  vector problem statement, stationary solution of the Navier-Stokes equation is assumed to be given; so, ${\bf v} (\br,t)$ is considered as a stochastic stationary vector field with known statistics.

In this case, for scales much smaller than Kolmogorov viscous scale, the equation (\ref{1}) can be solved,
and long-time approximations for
correlations of $\bB$ can be found \cite{Chertkov, epl18}. We will now make use of this solution and corresponding technique to calculate $$\tilde{\mathbf f} = (\bB \nabla) \bB $$ and its statistical moments.
$^1$\footnotetext[1]{The  gradient part of $\tilde{\bf f}$, which produces the right-hand side of (\ref{for-pressure}), can also be eliminated by renormalization of pressure; then the right-hand side in  (\ref{for-pressure}) would be zero. However, this procedure is laborious; instead, in what follows we show that the right-hand side of (\ref{for-pressure})      decreases to zero as well as both gradient and rotor components of $\tilde{\bf f}$. }

\vspace{0.4cm}
{\bf Passive vector dynamics. \ }
In the viscous range of scales the velocity field $\bf v$ is linear:
$$
v_i = A_{ij} {r_j} \ , \qquad A_{ii}=0 \ ,
$$
the strain tensor $A_{ij}(t)$ is an isotropic random matrix process with correlation time $\tau_c \ll t$.
The traceless condition is the result of incompressibility.

We introduce the evolution matrix $\mathbf{Q}$ that obeys
\begin{equation} \label{Qdef}
\dot{\mathbf{Q}} = - \mathbf{Q} \mathbf{A}   \ , \qquad  \mathbf{Q}(0)=\mathbf{I}
\end{equation}

The solution to the linear equation (\ref{1}) can be easily found by the Fourier transform (combined with change of variables  $\br' = \mathbf{Q} \br$):
$$
B_m (\br, t) =  Q^{-1}_{mn} \int  e^{i\bp\mathbf{Q}\br} B_n(\bp,0)
 e^{-\varkappa p_{i}p_j \int  (\mathbf{Q}\mathbf{Q}^T)_{ij}(t')dt'} d\bp
$$
Then
\begin{equation} \label{BgradB}
\begin{array}{l}
\tilde f_m (\br, t)  = \int  e^{i \bp \mathbf{Q} \br} B_n(\bp,0) e^{-\varkappa p_{i}p_j \int  (\mathbf{Q} \mathbf{Q}^T)_{ij}(t')dt'} d\mathbf{p} \\
\times Q^{-1}_{mk} \int i p'_n e^{i \bp' \mathbf{Q} \br} B_k(\bp',0) e^{-\varkappa p'_{a}p'_b \int  (\mathbf{Q} \mathbf{Q}^T)_{ab}(t')dt'} d\mathbf{p}'
\end{array}
\end{equation}

To get statistical moments of this quantity, we have to average over initial magnetic field and over $A(t)$.
Because of homogeneity of the flow, in what follows we restrict our consideration with $\br =0$.

\vspace{0.4cm}
{\bf Homogeneous initial conditions. \ }
We now have to specify the initial conditions for magnetic field.
 For simplicity,  $\bB (\bp, 0)$  is assumed to be Gaussian. In statistically homogenous case the initial pair correlator depends only on the difference $\br-\br'$, so in terms of Fourier transform we get
\begin{equation} \label{initialhom}
 \langle B_n(\bp,0) B_m (\bp',0) \rangle =  \delta(\bp+\bp') N(p) e^{-p^2 l^2}  \Pi_{mn} \ ,
 \end{equation}
$$
  \Pi_{mn} =  p^2 \delta_{mn}  - p_m p_n
$$
The multiplier $N(p)$ is arbitrary and does not affect the result; following \cite{Chertkov}, we take $N=1$.
According to the Wick theorem, the forth- and higher-order correlators are combined from products of double correlators with  all possible combinations of indices.

To average the square of (\ref{BgradB}), we make use of the Wick theorem; one of three summands is zero since
 $p'_n \left \langle B_n (\bp,0) B_k(\bp',0)\right \rangle \propto p'_n \Pi_{nk}(\bp') =0$.
Substituting (\ref{initialhom})
we get
\begin{equation} \label{inc-average}
\begin{array}{c}
\phantom{\int \limits_{a}}
\langle \tilde{ f}^2 
\rangle_{i.c.}  =
 (\mathbf{Q} \mathbf{Q}^T )^{-1}_{mk} \int d \bp d \bp'\,  e^{\displaystyle -(p_i p_j+p'_i p'_j) \tilde{D}_{ij}} \\
\times \left[ \left( \bp\Pi(\bp') \right)_k \left( \bp' \Pi(\bp) \right)_m + \Pi_{mk}(\bp') p'_n p'_l \Pi_{nl}(\bp) \right]
\end{array}
\end{equation}
where
\begin{equation} \label{D-stat}
\tilde{D}_{ij} = 2 \varkappa \int  (\mathbf{Q}\mathbf{Q}^T)^{-1} _{ij}(t')dt' + l^2 \delta_{ij}
\end{equation}
From asymmetry of the integrand it follows
$\int p_k \Pi_{nj}(\bp)e^{(\dots)}=0$,  so the first term vanishes.

\vspace{0.4cm}
{\bf Polar decomposition for the evolution matrix. \ }
Now, we make use of the polar decomposition for the evolution matrix:$^{1}$\footnotetext[1]{Note that $\zeta_i$ differ by the sign from $z_i$ used, e.g., in \cite{epl18}. }
$$
\mathbf{Q} = {\bf s d R } \  , \qquad {\bf s},{\bf R} \in SO(3) , \qquad  {\bf d} = \mbox{diag} \{ e^{-\zeta_i t} \} \
$$
The incompressibility condition implies $\det {\bf d}=1$, $\sum \zeta_i = 0$.
It is well known \cite{Let} that the long-time asymptotic behavior of these three components 
is quite different: as $\mathbf{Q}$ obeys Eq. (\ref{Qdef}),  ${\bf s}(t)$ stabilizes at some random value that depends on the realization of the process; $\zeta_i$ tend (with unitary probability) to  the limits $\lambda_i$,  $\lambda_1 \ge \lambda_2 \ge \lambda_3$, the set of $\lambda_i$ is the Lyapunov spectrum \cite{Oseledets};
and ${\bf R}(t)$ remains rotating randomly.
We note that, since
$\mathbf{Q}\mathbf{Q}^T = {\bf s} \mathbf{d}^2{\bf s}^T$,
the matrix $\bf R$ vanishes in the (\ref{inc-average}) and all correlators.

To eliminate the rotation matrices, we change the integration  variables for $\tilde{\bp} = \bp {\bf s}$, $ \tilde{\bp}'=\bp' {\bf s} $ (and omit tildes again, for brevity). With account of
 ${\bf s}^T \Pi (\bp') {\bf s} = \Pi ( \tilde{\bp}')$, $\bp'\Pi(\bp) \bp' = \tilde{\bp}'\Pi(\tilde{\bp}) \tilde{\bp}'$  we get
\begin{equation} \label{star-stat2}
\langle B^2 (t) \rangle_{i.c.} =   \int d^{-2}_{mn}   \Pi_{mn}(\bp')  e^{\displaystyle -p'_i p'_j D_{ij}} d \bp'\ ,
\end{equation}

\begin{equation}\label{force-ic-average}
\begin{array}{rl}
\langle \tilde{f}^2(t)\rangle_{i.c.} &= \int d{\bp}   \Pi_{mn} ({\bp})  e^{\displaystyle -p_i p_j {D}_{ij}} \\
&\times  \int d{\bp}'   p'_m   p'_n   \Pi_{kl} ({\bp}')  d^{-2}_{kl}   e^{\displaystyle -p'_i p'_j {D}_{ij}}
\end{array}
\end{equation}
where
\begin{equation} \label{defD}
\begin{array}{c}
\displaystyle
D_{ij} = l^2 \delta_{ij} + 2\varkappa \int \left( {\bf s}^T(t) {\bf s}(t') {\bf d}^2 {\bf s}^T(t') {\bf s}(t) \right) _{ij} dt'
\\
= \left( l^2 + \frac{\varkappa}{\lambda_i} e^{-2 \zeta_i t} \theta(-\zeta_i)  \right) \delta_{ij}  + \varkappa \mathbf{M}[{\bf s}(t)] \ , \quad   M_{ij} =O(1)
\end{array}
\end{equation}
Here we pick out the  part of $D_{ij}$ that grows exponentially: in these terms, 
the products of $\bf s$ cancel because of its stabilization. The rest terms do not grow; they are all gathered in the matrix $\mathbf{M}$, and we do not need them in what follows.  

\vspace{0.3cm}

We now introduce
$$
J_{mn} =  \int {p_m p_n} e^{\displaystyle -p_i D_{ij} p_j }  d \bp  = \delta_{mn} J_m \ ,
$$
$$
K_{mn} =  \int {p_m^2 p_n^2} e^{\displaystyle -p_i  D_{ij} p_j} d \bp \ ;
$$
with account of $D_{ij}=D_i \delta_{ij}$ (no summation), we have
\begin{equation}   \label{J}
J_m = \frac{I_0}{2D_m} \ , \  K_{mn}= \frac{I_0}{4 D_m D_n} \left(1+2 \delta_{mn}\right) \ ,
\  I_0=\frac {\pi^{3/2}}{ \prod_j\sqrt{D_j} }
\end{equation}

From (\ref{star-stat2}) we get
\begin{equation}   \label{star-stat3}
\begin{array}{rcl}
\left\langle B^2 (t) \right\rangle_{i.c.}
&=& \left( e^{2\zeta t} \right)_{ij} \left( \delta_{mn}\delta_{ij}  -  \delta_{im}\delta_{jn} \right) J_{mn} \\
 &=&    \sum \limits_i  e^{2 \zeta_i t}  \left( tr J - J_i \right)
 \end{array}
\end{equation}

From (\ref{force-ic-average}) we then have
\begin{equation}   \label{f-JK}
\left \langle \tilde{f}^2(t)  \right \rangle_{i.c.} =
\sum \limits_n \left[
 \sum_{a\ne n} J_a
 \cdot  \sum \limits_k \left( e^{2\zeta_k t}  \sum \limits_{b\ne k} K_{nb} 
 \right)
\right]
\end{equation}

As a vortex tube lengthens, its transverse size decreases; while it remains bigger than the diffusivity scale
$r_d \sim \sqrt{\varkappa/ \lambda_1}$, one can use the approximation of ideal conductive medium; diffusivity is negligible.
In (\ref{defD}) this corresponds to $D_{ij} \simeq l^2 \delta_{ij}$. In this approximation we get
$
 \langle \tilde{f}^2 (t)\rangle_{i.c.} \sim e^{2\zeta_1 t}
$.
This means exponential increase for the realizations with $\zeta_1>0$.
 However, as $\zeta_1$ is big enough, the integral in (\ref{defD}) becomes comparable with $l^2$  (since $\sum \zeta_i = 0$, i.e., $\zeta_3\sim -\zeta_1$), and the ideal conductor approximation is no more valid.

In the long-time evolution we have%
$^1$\footnotetext[1]{Hereafter we assume the ordering of $\zeta_i$ same as the ordering of $\lambda_i$. This is not necessarily so in an arbitrary realization, but the realizations with different ordering have exponentially small probability and  do not contribute to the averages. One can check this by calculation  their contributions by means of the technics developed in \cite{PRE2020}. }
 \,  $D_{ij} \simeq \mathrm{diag} \{ l^2, \frac{\varkappa}{\lambda_2} e^{-2 \zeta_2 t}, \frac{\varkappa}{\lambda_3} e^{-2 \zeta_3 t} \}$ for $\zeta_2<0$ and $D_{ij} \simeq \mathrm{diag} \{ l^2, l^2,  \frac{\varkappa}{\lambda_3} e^{-2 \zeta_3 t}  \}$ for $\zeta_2 \ge 0$. The two cases correspond to the 'filament' and 'pancake' scenarii of fluid element's deformation.

In the case $\zeta_2<0$ we have $J_1\gg J_2 \gg J_3$ and $\sum_a J_a - J_n \sim(e^{2\zeta_2 t}, l^{-2}, l^{-2})I_0$, the main contribution to (\ref{f-JK}) comes from $k=1,n={1,2}$;
for $\zeta_2 > 0$ one obtains $J_1\sim J_2 \gg J_3$ and all $\sum_a J_a - J_n \sim l^{-2} I_0$ for any $n$.
Substituting (\ref{J}) in (\ref{star-stat3}) and (\ref{f-JK}) we get
\begin{equation}   \label{B2 homogen}
\left\langle {B}^2 (t) \right\rangle_{i.c.} \propto \left\{ \begin{array}{ll} e^{(\zeta_1-\zeta_2) t} & \ \mbox{if} \  \zeta_2 \ge  0 \  \\
e^{(\zeta_2-\zeta_3) t} & \ \mbox{if} \  \zeta_2<0 \ \end{array} \right. \ ,
\end{equation}
\begin{equation}   \label{f-homogen}
\left\langle \tilde{ f}^2 (t) \right\rangle_{i.c.} \propto
e^{2 \alpha \zeta_2 t} \ , \ \
\alpha=
\left\{ \begin{array}{ll}
 -1 & \ \mbox{if} \  \zeta_2 \ge 0 \ , \\
  2 & \ \mbox{if} \  \zeta_2<0 \ \end{array} \right.
\end{equation}
We see that $B^2$ increases exponentially for all considered realizations,  and
$\tilde{ f}^2$ decreases independently of the sign of $\zeta_2$; thus,
\begin{equation}   \label{B2 homogen-log}
  \frac 1t \left\langle \ln \left\langle B^2 (t) \right\rangle_{i.c.}\right\rangle_A = \left\{ \begin{array}{ll} \lambda_1-\lambda_2 & \ \mbox{if} \  \lambda_2 \ge 0 \ , \\
\lambda_2-\lambda_3 & \ \mbox{if} \  \lambda_2<0  \end{array}  \right. \ ;
\end{equation}
\begin{equation}   \label{f-log}
  \frac 1t \left\langle \ln \left\langle \tilde{ f}^2 (t) \right\rangle_{i.c.}\right\rangle_A =
\lambda_2-3|\lambda_2|
  \end{equation}
 Here the outer brackets denote the average over the realizations of $A_{ij}$.

 This result shows that the non-gradient component of the Lorentz force does not grow exponentially, although~$B^2$ does.
  This does not mean that $\nabla \bB$ itself is small. To the contrary, e.g., for the second term in (\ref{force}) 
  one can by means of the same technique find
   $$
    \frac 1t \left\langle \ln \left\langle  \left(\nabla B^2 \right)^2 \right\rangle_{i.c.}\right\rangle_A = \left\{ \begin{array}{ll}
4 \lambda_1 & \ \mbox{if} \  \lambda_2 \ge 0 \ , \\
2 \lambda_2 -    4  \lambda_3  & \ \mbox{if} \  \lambda_2<0
  \end{array}  \right.
  $$
So, this effective addition to the pressure also increases. However, the right-hand side of (\ref{for-pressure}), which is $\nabla \cdot \tilde{\bf f}$, satisfies
 $$
    \frac 1t \left\langle \ln \left\langle  \left(\d_i B_j \d_j B_i \right)^2 \right\rangle_{i.c.}\right\rangle_A = \left\{ \begin{array}{ll}
2 \lambda_3 & \ \mbox{if} \  \lambda_2 \ge 0 \ , \\
4 \lambda_2 -    2  \lambda_1  & \ \mbox{if} \  \lambda_2<0
  \end{array}  \right.
  $$
Thus, the long-time asymptotic equation for the effective pressure does not contain magnetic field; the right-hand side of (\ref{for-pressure}) decreases to zero.

\vspace{0.4cm}
{\bf Statistical moments of $\tilde{\bf f}$. \ }
To calculate  statistical moments of (\ref{f-homogen}) we have to average its powers over all realizations of $A$; this is, over different realizations of $\zeta_2$, since it is - to logarithmic accuracy - the only functional of $A(t)$ in  (\ref{f-homogen}).
To this purpose, it is convenient to use the formalism of Cramer function \cite{klassiki-large-dev}.
By its definition, in the limit $t\to \infty$ the probability density $\mathcal{P}(\zeta_2)$ is
$$
\mathcal{P}(\zeta_2)=  e^{-tS(\zeta_2 - \lambda_2)}
$$
where the Cramer function $S(y)$ satisfies $S(0)=0$, $S'(0)=0$, $S''>0$ $\forall y$.

So, the $n$-th order moment of $\tilde{\bf f}$ can be found as
\begin{align}
\notag
& \left\langle  \tilde{f}^n(t)  \right\rangle
=\int d\zeta_2    e^{G(\zeta_2)t} \ ,
 \\& G= n \alpha \zeta_2 - S(\zeta_2 -\lambda_2)
\end{align}
We take the integral  by means of the saddle point method. The derivative of the exponent is
\begin{equation*}
d G / d \zeta_2 = n \alpha (\zeta_2) - S'(\zeta_2- \lambda_2) \ , \ \zeta_2 \ne 0
\end{equation*}
 Consider first the case $\lambda_2>0$; then $S'(0)<0$. Since $S'$ is  monotonic function, we find that
 the point $\zeta_2^*$ where $G$ reaches its maximum satisfies
\begin{align*}
 S' (\zeta_2^*-\lambda_2)& = - n  \ \  \mbox{if} \  n<-S'(0) \ ; \\ \zeta_2^* &= 0 \   \mbox{if} \  n \ge -S'(0)
\end{align*}
As $n$ increases,  $\zeta_2^*$ decreases to zero; for $n\ge -S'(0)$,  the maximum of $G$ is situated at the point of discontinuity of $\alpha$, $\zeta_2^*=0$.

\begin{figure}
\center{\includegraphics[width=0.75\linewidth]{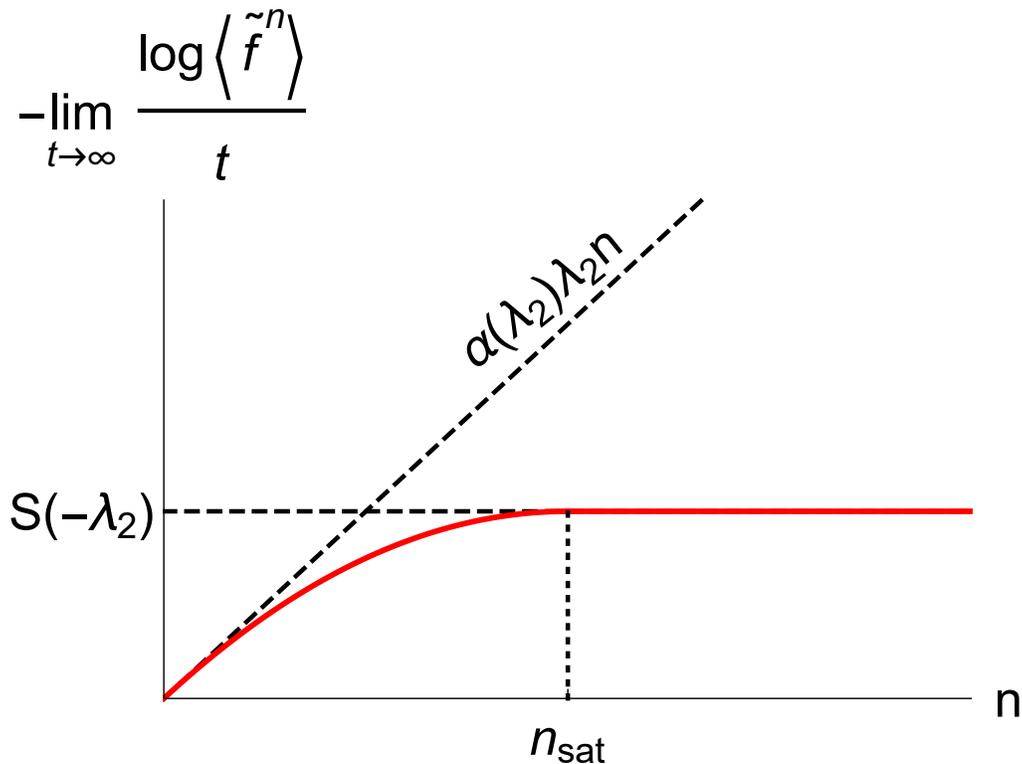}}
\caption{Decrements of the n-th moments of $\tilde{f}$.}
\end{figure}

The case $\lambda_2<0$ is almost symmetric; now $S'(0)>0$, and $\zeta_2^* \le 0$.  Similarly to the first case, $\zeta_2^*$ begins with $\lambda_2$ as $n=0$;  as $n$ increases, $\zeta_2^*$ approaches zero and stays equal to zero for  $n \ge S'(0)/2 $.

Summarizing, for all signs of $\lambda_2$ we get
~
\begin{align*}
 S' (\zeta_2^*-\lambda_2) &= \alpha n   \ \  \mbox{if} \  n<S'(0)/\alpha(\lambda_2) \ ;
\\
\zeta_2^* = 0& \ , \ \   \mbox{if} \  n \ge \frac{S'(0)}{\alpha(\lambda_2)}
\\
\alpha (\lambda_2) &= 1/2 - 3/2 \; \rm{sign} (\lambda_2)
\end{align*}

In all cases, to logarithmic accuracy we have
\begin{equation}
 \left\langle \tilde{f}^n (t) \right\rangle  \propto  e^{G(\zeta_2^*) t}
\end{equation}

We note that $G(\zeta_2^*)$ remains negative for any $n > 0$ and $\lambda_2 \ne 0$. So, {\it all} statistical moments of $\tilde{\bf f}$ decrease as functions of time.
The decrement $G$ has linear asymptote,  $G \simeq \alpha(\lambda_2) \lambda_2 n$ for $n\ll 1$ (see Figure 1). However,
this decrease is strongly intermittent: the decrements saturate beginning with some $n$, $G(n \ge -S'(0)/\alpha(\lambda_2)) = S(-\lambda_2)$. An analogous effect of saturation of damping decrements was found in \cite{BF} for advected passive scalar, and in \cite{epl18} for localized perturbations of small-scale advected magnetic field.


In the exceptional case $\lambda_2=0$ the decrease is not exponential ($G(\zeta_2^*)=0$) but a power law;
one can check that for Gaussian probability distribution of $\zeta_2$, $\mathcal{P}_G \propto \sqrt{t} e^{-\zeta_2^2 t/2D}$,
the statistical moments of $\tilde {\bf f}$ are proportional to $1/\sqrt{t}$ for any $n$.

The values of $S'(0)$, $S(0)$, as well as the whole shape of $S$, are determined by the statistics of velocity gradients. One can show (by means of the technics developed in  \cite{PRE2020, epl2020, JOSS1}) that for isotropic $A_{ij}(t)$,  possible value of
$S'(0)$
is restricted by the boundaries $|S'(0)|< 3/2$.
 Thus, for these processes saturation of $\langle \tilde{f}^n \rangle$ happens already at $n<3/2$,  so for all integer $n \ge 2$ the $n$-order moments decrease with the same exponent.

In conclusion, we recall that the observed decrease of the 'effective'   part of the Lorentz force is related closely to magnetic diffusivity of the flow.
The 'ideal conductor' approximation (zero diffusivity) was first analyzed in \cite{Kleeorin}; it was shown that
the Lorentz force increases exponentially. Now we see that even small diffusivity is essential: it
 makes the increase of magnetic field slower (although still exponential), and all the moments  $ \langle  \tilde{ f}^n  \rangle  $ decrease exponentially. So, the Lorentz force remains small for any finite Prandtl number.

So, the nonlinear feedback of magnetic field on velocity dynamics never happens in the classical frame of viscous statistically homogenous flow.

The authors thank 
Prof. A.V. Gurevich for his kind attention to their work.
This work was supported by the RAS program 'Extreme phenomena and coherent structures in nonlinear physics'.

\end{document}